\documentclass[a4paper,fleqn]{cas-sc}
\usepackage{natbib}
\usepackage{longtable}
\usepackage{pdflscape}
\def\tsc#1{\csdef{#1}{\textsc{\lowercase{#1}}\xspace}}
\tsc{WGM}
\tsc{QE}
\begin{document}
\let\WriteBookmarks\relax
\def\floatpagepagefraction{1}
\def\textpagefraction{.001}

\shorttitle{ChaI Association}    

\shortauthors{K. Neumannova et al.}  

\title [mode = title]{Lightcurves of stars in the Chamaeleon I Association}  

\tnotemark[1] 

\tnotetext[1]{} 


\author[1]{ K.~Neumannov{\'a}}%[<options>]

\affiliation[1]{organization={Department of Theoretical Physics and Astrophysics - Masaryk University},
            addressline={Kotl\'a\v{r}sk\'a 2}, 
            city={Brno},
%          citysep={}, % Uncomment if no comma needed between city and postcode
            %postcode={}, 
            %state={Czechia}
            country={Czechia}
            }

\author[2]{L.~Kueß}%[]

% Footnote of the second author
\fnmark[2]

\affiliation[2]{organization={Department of Astrophysics - Vienna University},
            addressline={T{\"u}rkenschanzstraße 17}, 
            city={Vienna},
%          citysep={}, % Uncomment if no comma needed between city and postcode
            postcode={}, 
            %state={Austria}
            country={Austria}
            }

\author[1]{E.~Paunzen}%[]

% Footnote of the third author
\fnmark[3]

\author[3,4]{K.~Bernhard}%[]

% Footnote of the third author
\fnmark[4]

\affiliation[3]{organization={Bundesdeutsche Arbeitsgemeinschaft f{\"u}r Ver{\"a}nderliche Sterne e.V. (BAV)},
            addressline={Munsterdamm 90}, 
            city={Berlin},
%          citysep={}, % Uncomment if no comma needed between city and postcode
            %postcode={}, 
            %state={Germany}
            country={Germany}
            }

\affiliation[4]{organization={American Association of Variable Star Observers (AAVSO)},
            addressline={49 Bay State Road}, 
            city={Cambridge},
           %citysep={}, % Uncomment if no comma needed between city and postcode
            %postcode={}, 
            state={Massachusetts},
            country={USA}
            }

\fntext[1]{}


\begin{abstract}
Star-forming regions are essential for studying very young stellar objects of various masses. They still contain a significant amount of dust and gas. We present a study of light curves of stars in the field of the Chamaeleon I association. We use automatic spectral classification with MKCLASS to identify the spectral types of the stars in the field with a light curve from the NEOWISE and Gaia surveys. The light curves are analysed using the software Peranso and astropy. We also used VSX to identify the variability type. Based on astrometry, we have identified 92 stars, 73 of which are members of the association. We received light curves for 55 stars from the Gaia survey and for 69 stars from the ALLWISE/NEOWISE survey. For 28 of them, it was possible to determine the types of variables, mostly T Tauri and Orion variables. The spectral types of the members are mostly cooler M-type stars, with one being a possible chemically peculiar (CP) star. The non-members associated with light curve measurements include spectral types A-G with one CP candidate.

%\nocite{*}%% Remove this line from your manuscript.
\end{abstract}


%\item 
%\item 
%\item 

% Keywords


\maketitle

% Main text
\section{Introduction}\label{}

The Chamaeleon I (Cha I) association is one of the closest star-forming regions to the Sun. It has been well studied in the past to determine membership probabilities and ages to the association \citep[e.g.][]{2004ApJ...602..816L, 2007ApJS..173..104L, 2021A&A...650A..48K}. With an age of $\sim$ 2Myr \citep{2007ApJS..173..104L}, it is also one of the youngest stellar aggregates known and thus is an interesting target for observing star formation.
 
 Pre-main-sequence (PMS) stars provide essential insights into star formation and the processes happening during this phase. Most observations and analyses can only be done using infrared (IR) because the optical wavelengths are blocked by the dust and clouds surrounding star-forming regions \citep{2003ARA&A..41...57L}. However, some studies can be performed using the optical wavelengths, given the cloud is sparse enough to let through at least a portion of the light emitted by the young stars. Not only is it challenging to observe stars in the PMS phase because of the dust, but also because of the short time the star remains in this phase.
 
 Light curves (photometrically time series) are an essential tool in modern astronomy. In addition to detecting and classifying new variable sources, it can also help determine the period of change in brightness over time, the mass, size and temperature of a star, and give us insight into the inner workings of stars. This process is called asteroseismology for pulsating stars. While it has been successful for all kinds of pulsating variable stars on the main sequence and giant stars \citep{2021RvMP...93a5001A}.
 It also has been proven a valuable tool to infer the evolution of PMS sources \citep{2014Sci...345..550Z}.
 
 However, while time series analysis for multiple PMS stars in other regions \citep[e.g.][]{2000AJ....120..349H,2021ApJ...921..165S} and even a study of variability in the infrared in Cha I \citep{2016ApJ...833..104F} have been conducted, no such attempts have been made for Cha I in the optical, possibly because the association is still primarily embedded in the accompanied molecular cloud (MC).
 
 In this work, we investigate periods in light curves in the optical region, members of the association and their spectral types. Sections 2 and 3 describe procedures for obtaining photometric and spectroscopic data and for identifying members of the association, section 4 describes the frequency analysis and spectral classification and determination of period, while in section 5 we study the ages of association. In sections 6 and 7, we present our results and give a conclusion in section 8.
 
 \begin{figure*}
    \begin{center}
	\begin{tabular}{cc}
    {\includegraphics[width=0.45\textwidth]{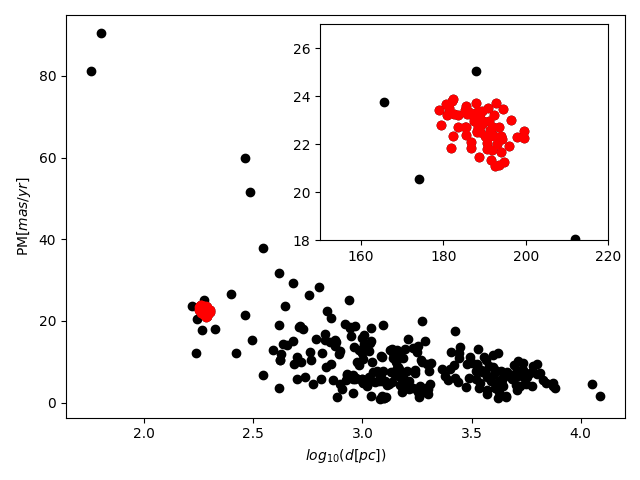}} \quad &
	{\includegraphics[width=0.45\textwidth]{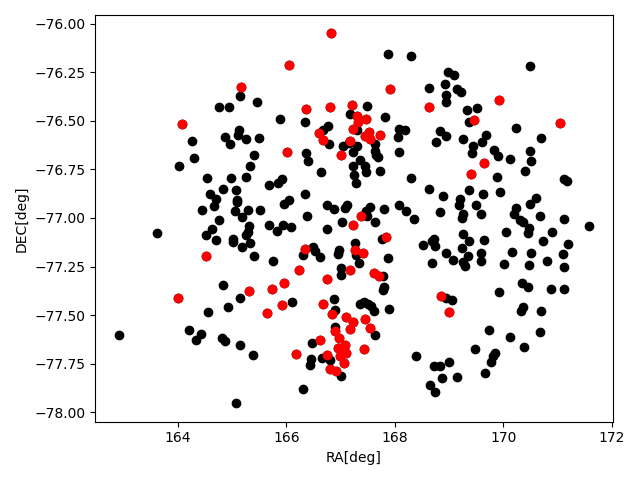}} \\
	\end{tabular}
	\end{center}
    \caption{Observed stars (black) and our members of the ChaI association (red). \textit{Left}: distance-proper motion space to determine the members as described in the text. \textit{Right}: sky distribution of the observed field.}
    \label{fig:targets}
\end{figure*}

\section{Target selection and published membership probabilities}

The Cha I association was observed in May 2009 with the AAOmega spectrograph \citep{2004SPIE.5492..410S} at the Anglo-Australian Telescope (AAT). 
The instrument consists of a fibre plate with 400 fibres and has a field of view of 2 degrees. 
This makes studying a relatively large portion of the sky useful. 
Depending on the specific configuration, the instrument can observe the blue and red spectrum in the wavelength 
range from 3600\AA\, to 9000\AA\, and a spectral resolution between 1300 and 10000. For our sample, the blue grid 580V (R$\sim$1300) and the red grid 2000R (R$\sim$8000) were used.
The targets were selected by taking the Field-of-View (FOV) of the instrument and centering it on the coordinates $\alpha$ = 11h 00m 00s, $\delta$ = -77d 00m 00s. 
All stars in a radius of two degrees around that point down to a magnitude $V$\,=\,15\,mag were selected as targets for the observations, with special interest in the known members of the association at that time. Notice that the data from the $Gaia$ mission were not available. The best candidates were selected by the catalogue of \citet{2007ApJS..173..104L}. 
The here presented sample of members is limited according to the region of the sky and the apparent
visual magnitude.
This resulted in 325 observed stars (members and non-members) in addition to several sky fibers for data reduction.\\
From the 325 stars observed, we took the astrometry from Gaia DR3 \citep{2016A&A...595A...1G,2022arXiv220800211G} 
to determine which ones belong to the association. To not throw out too many possible members in an already limited investigation, no further quality cuts (e.g. a threshold on $\varpi/\sigma_\varpi$ were performed. Of course, if taken such an approach, stars with poor astrometric solutions would be removed from the sample.\\
The stars with a distance between 170-210 pc (taken from \cite{2021AJ....161..147B} and a proper motion of 20-25 mas/yr based on the distribution of distance and proper motion (Fig. \ref{fig:targets}) were assumed to be members of this association. The values in proper motion and parallax were taken from looking at the distribution of the stars in Fig. \ref{fig:targets}, where one can see a relatively tight cluster inside the box defined by $d\in[170\textnormal{pc},210\textnormal{pc}]$ and $\mu \in [20\textnormal{mas}, 25\textnormal{mas}]$. That process left us with 67 stars in the association (See Fig. \ref{fig:targets}).

Additionally, we also determined memberships using the \textit{Hierarchical Density-Based Spatial Clustering of Applications with Noise} (HDBSCAN, \cite{inproceedings, McInnes2017}) on an input dataset containing euclidean $X,Y,Z$ distances (in pc) and galactic transversal velocities $v_{t,l}, v_{t,b}$ (in $km\,s^{-1}$). This resulted in an updated list of 73 sources with a membership probability $p>0.5$. Those stars can be seen in Table \ref{tab:members_hdbscan}. This is also the list of stars that was used in further light curve analysis.

\begin{figure}
    \centering
    \includegraphics[width = 0.65\textwidth]{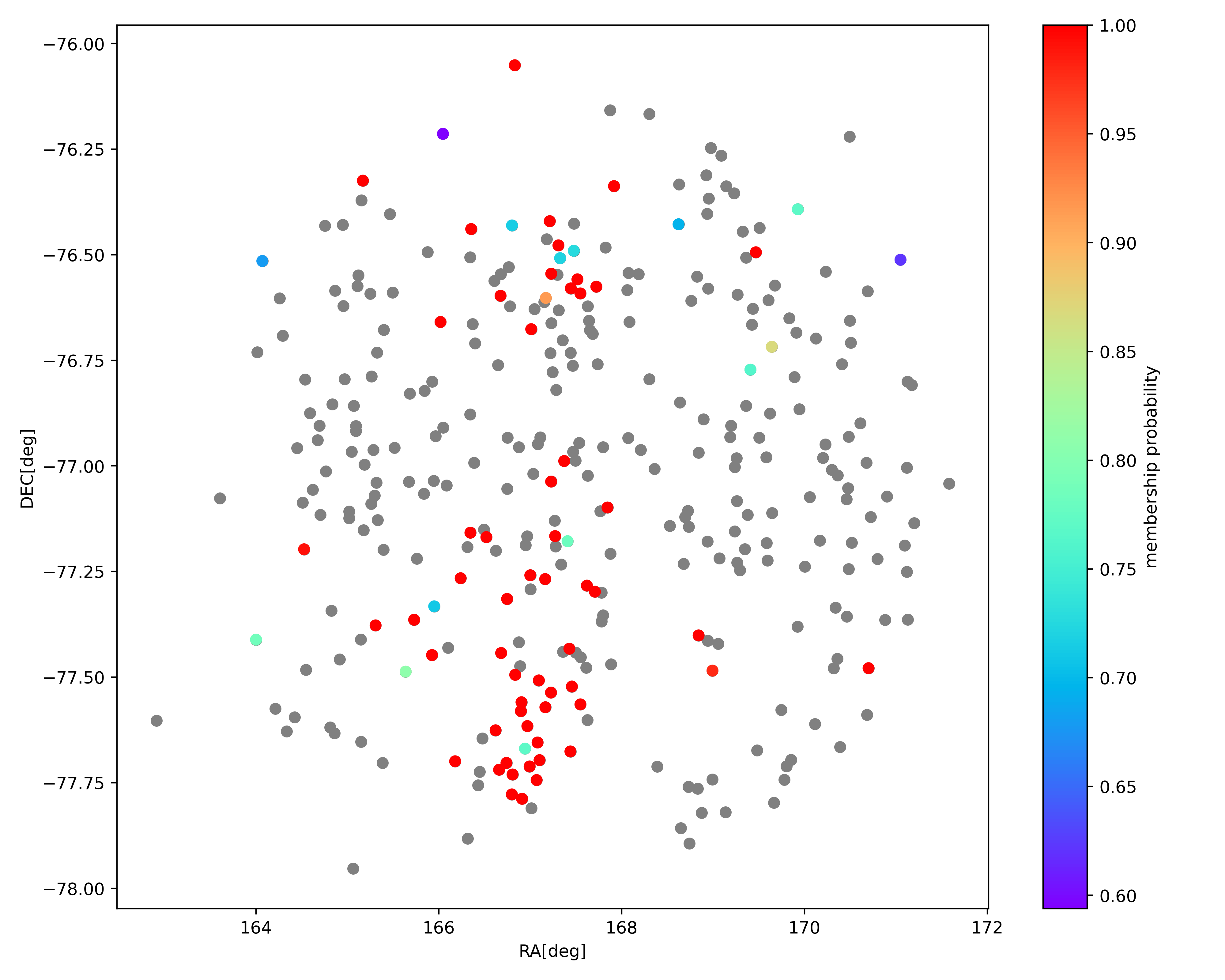}
    \caption{Members of the association determined by HDBSCAN. The stars are colour-coded by membership probability. The grey points are foreground and background sources.}
    \label{fig:enter-label}
\end{figure}

We note that our selection of membership is solely based on astrometric properties of the sample and that we did not use other criteria on the selection of members. One could also use the presence of emission lines (primarily in H$\alpha$) or the measurement of equivalent widths of lithium, which would indicate a (very) young age. However, we did not use these methods because of the rather low quality of our spectral data (see section \ref{spectypes}. So, our best chance in accurate determination of membership to the association was the aforementioned astrometry.

As the last step, the sample was also compared to the literature to check the membership. We used the catalogues of \cite{2007ApJS..173..104L}, which lists 226 members of the association, and \cite{2020A&A...643A..71G} with a list of 713 members. The former includes all of our members from the astrometric selection based on distance and proper motion and the latter had 42 ($\sim$ 63\%) of our members included. Those discrepancies arise mainly from the membership criteria used in the respective works. While we only use astrometry, i.e. parallax (distance) and proper motions, \cite{2007ApJS..173..104L} used a combination of spectroscopic and photometric criteria and \cite{2020A&A...643A..71G} worked with a combination of astrometry, mainly distance and radial velocity, Li-abundances, metallicity and surface gravity.

We also note a discrepancy between the distance to the association determined by us ($\sim$190\,pc) and the distances presented in the papers mentioned above. Both give a distance between 160 and 170\,pc, respectively. A comparison with other works \citep{2021A&A...650A..48K, 2020A&A...633A..51Z, 2018A&A...617L...4R} reveals similar distances to the one we defined.

\section{Light curves}

The IRSA (Infrared Science Archive)\footnote{\url{https://www.ipac.caltech.edu/}}, which includes several surveys and databases such as Gaia\citep[Gaia ESA Archive;][]{refId0}\footnote{\url{https://irsa.ipac.caltech.edu/Missions/gaia.html}} and WISE/NEOWISE/AllWISE \citep[Wide-field Infrared Survey Explorer;][]{Mainzer2014}\footnote{\url{https://irsa.ipac.caltech.edu/Missions/wise.html}} were used to obtain the light curves.

We used the equatorial coordinates  (RA and DEC) in the specified format as input parameters. We had the option to choose whether we only wanted data with exact matching values or whether we also considered the cone search radius. In our case, we performed a cone search with a radius of 5 arcseconds. During the subsequent processing, we consistently selected the data that most closely matched our specified colour values. The selected data that did not match precisely were manually checked to verify that they were not our target stars. 

The output from these catalogues includes information such as MJD (Modified Julian Date), magnitude (mag), and filter code. In addition, we obtained the minimum and maximum magnitudes of each star in G, BP, and RP bands. IRSA is an extensive catalogue that provides comprehensive tools for data analysis and display. Among other things, it allows us to display the time series for stars in different filters directly, obtain the light curve and manually determine the period, which displays the phase curve. This is therefore a quick check of the data.

We used the Python programming language to bulk process the available data in multiple colour filters for data processing. The light curves were then processed in Peranso \citep{2016AN....337..239P}.

\begin{figure*}[h]
    \begin{center}
	\begin{tabular}{cc}
    {\includegraphics[width=0.85\textwidth]{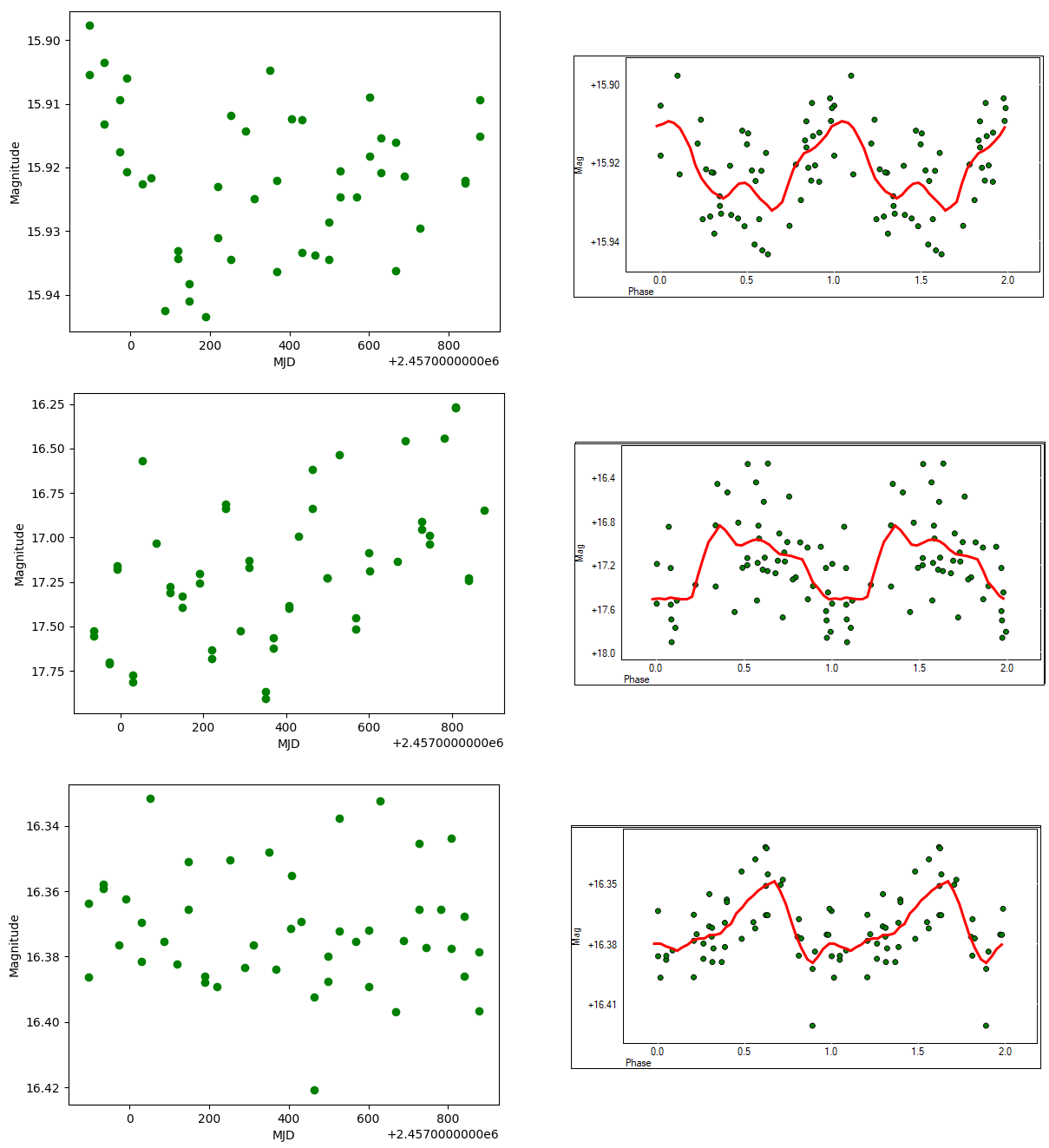}} 
	\end{tabular}
	\end{center}
    \caption{Light and phase curves of variable stars from the Gaia database from top to bottom; \textit{Left}: Light curve from Gaia database. \textit{Right}: Phase curve proceeds in Peranso. ID 5201341567397611008: period of 0.759 days. ID 5201335481425778560: period of 0.643 days. ID 5201127918543201664: period of 96.2 days.} 
    \label{fig:vse}
\end{figure*}

\section{Methods}

\subsection{Frequency analysis}

Lomb-Scargle, discrete Fourier transform (DFT) and ANOVA (analysis of variance) methods were used to determine the periods and phase curves in Peranso. The Lomb-Scargle method is a variant of the DFT, in which an unequally spaced time series is decomposed into
a linear combination of sinusoidal and cosinusoidal functions \citep{1976Ap&SS..39..447L,1982ApJ...263..835S}. The data are transformed
from the time to the frequency domain, particularly useful for analysing pulsating variable stars. From a statistical point of view, the resulting periodogram is related to the $\chi^2$ for a least-square fit of a single sinusoid to data, which can treat heteroscedastic measurement uncertainties. The underlying model is non-linear in frequency and the basis functions at different frequencies are not orthogonal. 
Here, the modifications of \citet{1986ApJ...302..757H} are incorporated. The power of the modified periodogram is normalized by the total variance of the data, yielding a better estimation of the frequency of the periodic signal.

To define whether a time series includes a periodic signal (or not), we followed the statistical approach published by \citet{2008MNRAS.385.1279B}. The false alarm probability (FAP) measures the likelihood that a data set with no signal would lead to a peak of a similar magnitude. We follow here the approach by \citet{2008MNRAS.385.1279B} who improved the analytic estimations based on extreme value theory.

Finally, we define when to consider a light curve not included as periodical signal for a given amplitude (or noise level). Using the results from the literature for (in particular) low-amplitude variables and non-variable stars of different spectral types \citep{2024A&A...687A.208P}, we decided to use a value (limit) of $\log FAP \geq -2$ for our subsequent analysis.

The ANOVA method combines Fourier analysis with periodic orthogonal polynomials to fit the data, making it effective in studying eclipsing variable stars of the \citep{Knote2019} type. Each of these methods identifies the dominant frequencies, allowing for an efficient determination of the periodicity.

The data obtained from these catalogues show slight variability in the quality of the light curves for most stars, with the data being burdened by irregularities in some cases. However, despite these factors, the identification of the light curve shape was unambiguous for most of the stars studied. None of the measured points deviated significantly from the others and therefore it was not necessary to proceed to their elimination. To automate the determination of the basic shape of the light curves, we used the Peranso tool with the fit mean curve function, which allowed an efficient approximation of the curves using the mean luminosity values as a function of phase. In cases where the data exhibited a higher degree of scatter or nonlinearity, we applied manual filtering methods to optimise the frequency spectrum. These methods involved manually adjusting the appropriate frequency, which allowed us to improve the accuracy of the resulting analysis and ensure the reliability of the determined light curve characteristics.

To validate the results obtained in Peranso, we used the Lomb-Scargle method implemented in the Astropy library in Python \citep{astropy:2022}. The resulting periods from Astropy differed only slightly from those obtained in Peranso, with the minimum identified period being 0.5 days in both cases. However, when we left this value undefined or set the default minimum period to 0.01 days, the resulting light curves were burdened by artefacts and aliasing, with the most significant frequency appearing close to this minimum value.

To determine whether our sample contains known variable stars, we compared our list with the Variable Star Index (VSX, \citep{2006SASS...25...47W}). We identified 29 stars matching our data, which we further analysed using the program UPSILoN (Automated Classification of Periodic Variable Stars) \citep{https://doi.org/10.48550/arxiv.1512.01611}. This machine learning tool analysed the time series of these objects and provided a type classification of variable stars. Unfortunately, in no case was classification confidence of more than 50 \% achieved, with most objects classified as Delta Scuti or RV Tauri stars.

We verified the results from the ANOVA and Lomb-Scargle method with several 
known T Tauri variables taken from the VSX (Table \ref{tab:vari_type}). The 
derived periods were identical within the error estimations.

\subsection{Spectral classification} \label{spectypes}

First, the spectra were reduced using the custom software \textit{2dfdr} \citep{2015ascl.soft05015A} and extracted into individual ASCII files via a Python script. Subsequently, they were flux calibrated using the spectrophotometric standard star $\theta$ Vir (HD 114330, HR 4963)\footnote{Calibrated spectrum obtained from \url{https://www.eso.org/sci/observing/tools/standards/spectra/hr4963.html}}, also observed by the same instrument.

 The spectra were then classified using version 1.07 of the MKCLASS code from \cite{2014AJ....147...80G}. This code takes the spectrum in the wavelength region between 3800\AA\, and 5600\AA\, and iteratively compares it to a library of standard spectra, resulting in a spectral classification that is comparable to the one made ``by eye''.
 
Some problems arise because the spectra show, for example, an emission feature, 
 likely due to the reference skyline at $\sim5570$\AA. Because we are interested in finding 
 emission lines as a characteristic of young stars, it was not a-priori eliminated. However, 
 the standard library \textit{libnor36} of MKCLASS got good results. 
 The other standard library, \textit{libr18}, could not initially determine spectral types. 
 This is caused by the fact that they were not rectified. 
 After rectification using a simple polynomial interpolation, 
 \texttt{libr18} worked properly. The result can be seen in 
 Table \ref{tab: spectral_types}.\\

\section{Results}

\subsection{Variability}

We had sufficient data to analyse 55 stars in the G band filter from the Gaia archive, and photometric data for 69 stars in the W1, W2, W3, and W4 filters from the AllWISE catalogue, covering association members. It turns out that the smallest magnitude difference is for Gaia DR3 5201341567397611008 with a value of 0.046 mag, while the largest difference in brightness at minimum versus maximum is for Gaia DR3 5201335481425778560 with a value of 1.6 mag, see top Fig. \ref{fig:vse} and middle Fig.\ref{fig:vse}. These stars are out of the normal trend, with most stars having a magnitude difference of about 0.3 mag.

Determining the frequency was time-consuming because we lacked sufficient regularly spaced data and had to employ multiple methods to ascertain it. This can result in frequencies with unclear values, significantly increasing the likelihood of inaccurately determining periods with a large margin of error. The average period of the brightness changes corresponds to approximately 6.3 days, while the longest period corresponds to Gaia DR3 5201127918543201664 with a period of 96.2 days see bottom of Fig.\ref{fig:vse}. 

All processed light and phase curves from the NEOWISE/AllWISE and Gaia archives can be found in the appendix of this article. The light curve figures from the Neowise/Allwise archive are named according to the object ID. For the Gaia archive data, we show both light curves and phase curves; the figure names correspond to the object ID, and the "phase" in the image name is followed by the object period in days, which was determined in Peranso using the Lomb-Scargle method.

We can also note that the phase curves are not smooth. Some T Tauri-type stars show similar phase curves. 
There could be several reasons for this. Unstable internal structure of the star in the PMS phase, when the resulting star is still forming. These stars are also surrounded by material that affects the radiation reaching the star. There is also a fluctuation in the accretion flux and a change in the accretion rate of material onto the star. Finally, these irregularities in the curve can be caused by spots that may appear on the star. According to \cite{Cody2018}, we would observe higher dips in the phase curve for older stars and for stars where we observe a disk with a larger inclination angle. This corresponds to the low amplitudes we observe belonging to very young stars.  They also hypothesize that stars with larger amplitudes may be affected by hot spots associated with intense accretion. In comparison, stars with smaller amplitudes and irregularities may be dominated by cool spots, where brightness changes may be more likely to result from a milder accretion process. These properties could have influenced the appearance of the phase curves we obtained.

\begin{figure*}[h]
    \begin{center}
	\begin{tabular}{cc}
    {\includegraphics[width=0.85\textwidth]{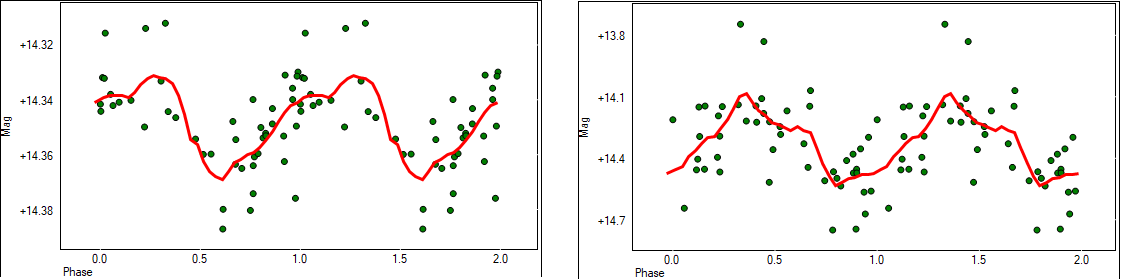}} 
	\end{tabular}
	\end{center}
    \caption{Phase curves of the variable stars of INS type. \textit{Left}: Phase curve of Gaia DR3 5201347064956072448 with period 5.1704 days. \textit{Right}: Phase curve of Gaia DR3 5201350019893311744 with period 0.6672 day.} 
    \label{fig:INS-both}
\end{figure*}

\begin{figure*}[h]
    \begin{center}
	\begin{tabular}{cc}
    {\includegraphics[width=0.85\textwidth]{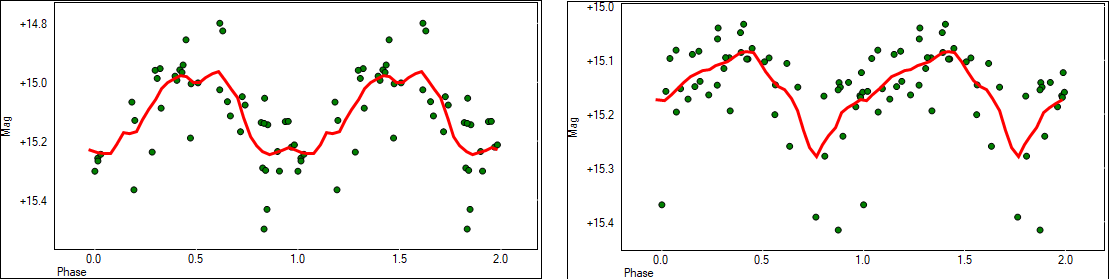}} 
	\end{tabular}
	\end{center}
    \caption{Phase curves of the variable stars of CTTS type. \textit{Left}: Phase curve of Gaia DR3 5201201139145692800 with period 1.1688 days. \textit{Right}: Phase curve of Gaia DR3 5201209351123255296 with period 0.6484 day.} 
    \label{fig:CTTS-both}
\end{figure*}

\begin{figure*}[h]
    \begin{center}
	\begin{tabular}{cc}
    {\includegraphics[width=0.85\textwidth]{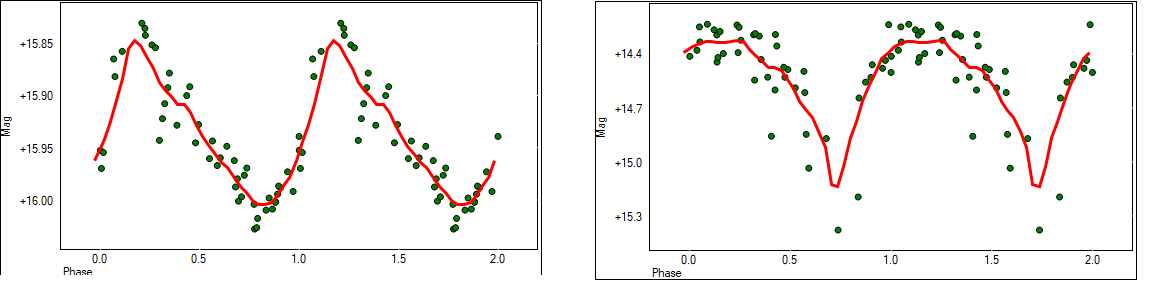}} 
	\end{tabular}
	\end{center}
    \caption{Phase curves of the variable stars of INT type. \textit{Left}: Phase curve of Gaia DR3 5201129705249611008 with period 4.8311 days. \textit{Right}: Phase curve of Gaia DR3 5201160525934988288 with period 0.7406 day.} 
    \label{fig:INT-both}
\end{figure*}

\begin{figure*}[h]
    \begin{center}
	\begin{tabular}{cc}
    {\includegraphics[width=0.85\textwidth]{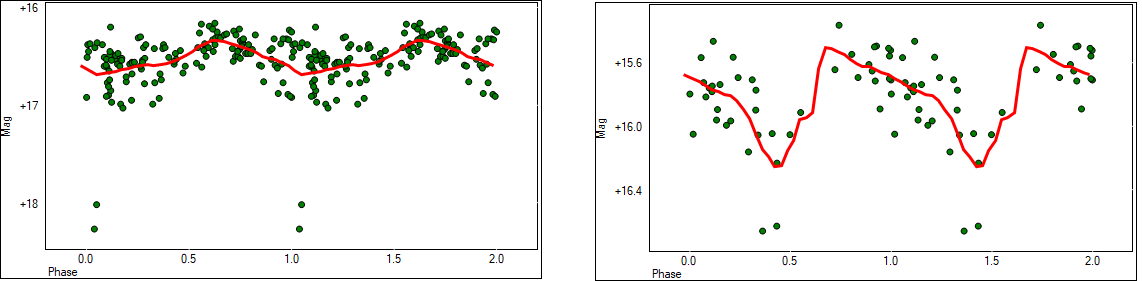}} 
	\end{tabular}
	\end{center}
    \caption{Phase curves of the variable stars of TTS and WTTS types. \textit{Left}: Phase curve of TTS type Gaia DR3 5201351428642607744 with period 6.7818 days. \textit{Right}: Phase curve of WTTS Gaia DR3 5201126441074447872 with period 1.055 days.} 
    \label{fig:TTS-WTTS-both}
\end{figure*}

\begin{figure*}[h]
    \begin{center}
	\begin{tabular}{cc}
    {\includegraphics[width=0.85\textwidth]{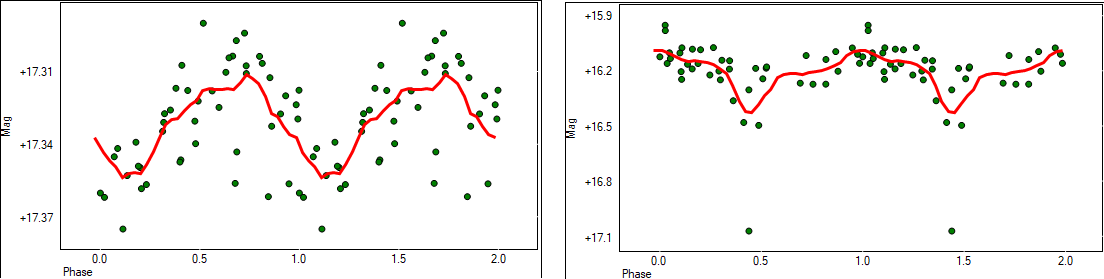}} 
	\end{tabular}
	\end{center}
    \caption{Phase curves of the variable stars of EB and SR types. \textit{Left}: Phase curve of EB type Gaia DR3 5201181313576638208 with period 0.8509 days. \textit{Right}: Phase curve of SR Gaia DR3 5201207736215474688 with period 1.4299 days.} 
    \label{fig:EB-SR-both}
\end{figure*}

From the VSX catalogue, we obtained variability types for 29 of our stars. As expected, most of them (13) fall into the Orion Variables of the T Tauri-type, which are young stars. Subsequently, the classical T Tauri and Orion variables are abundant. However, the program also identified one as an Eclipsing Binary and one as a Semi-regular variable. For these we rather assume that the types contained in the VSX files from different observers are partially incorrect or outdated for our stars. The full table can be found in the appendix, see table \ref{tab:vari_type}, where the notation is taken from \cite{2006SASS...25...47W}. For each type you can find in the figures light curves with specified periods, see figures; \ref{fig:INS-both}, \ref{fig:CTTS-both},\ref{fig:INT-both},\ref{fig:TTS-WTTS-both}, \ref{fig:EB-SR-both}.

\begin{figure}
    \centering
    \includegraphics[width =0.65 \columnwidth]{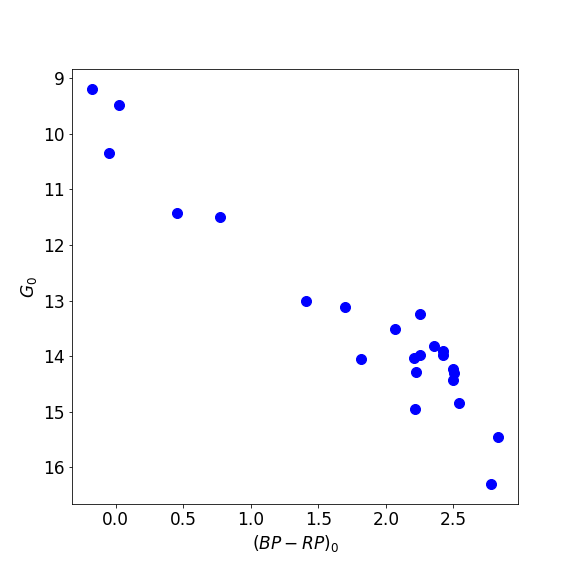}
    \caption{Dereddened colour-magnitude diagram of the Cha I association with the Gaia filters. The stars show linear dependence - The redder the object, the less bright it appears. As the redness increases, the extinction increases, as we would expect.}
    \label{fig:deredd.gaia}
\end{figure}

Using the Gaia catalogue we were able to obtain and subsequently process not only the minimum and maximum magnitudes of the stars but also the extinctions, which we then used for the dereddened CMD.
To calculate the dereddened values we used the relations $G_0 = G - AG$ and $(BP-RP)_0 = (BP-RP) - E(BP-RP)$ .
These data were available for about 1/3 of the stars. There is a clear linear dependence in Fig. \ref{fig:deredd.gaia}, with a higher colour value making the object less bright, as we would expect.

\begin{figure}
    \centering
    \includegraphics[width =0.65 \columnwidth]{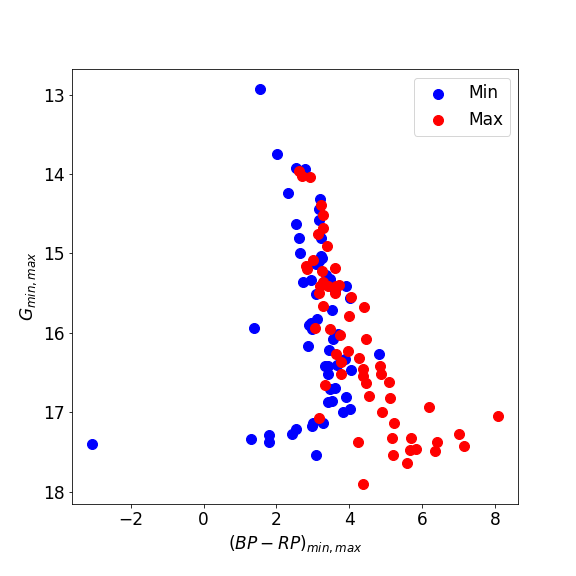}
    \caption{Colour - magnitude diagram of the Cha I association with the Gaia filters with minimum and maximum magnitudes for each star. We can notice how each star is shifted during its cycle.}
    \label{fig:color.gaia}
\end{figure}

A study of the magnitude variation during brightness changes due to variability can be seen in Fig. \ref{fig:color.gaia}, where the magnitude minima - i.e. the brightness maxima - are shown in blue and the reverse in red. These stars are redder when dimmer in $G$, thus variations in brightnesses are caused by changes in temperature, such as delta Scuti, RV Tauri and T Tauri variables, whose changes are caused by pulsation, accretion and/or magnetic activity.

The data with the specified variability type were checked against the Gaia DR3 variability catalogue to confirm that they are real variable stars. Approximately 66 \% of the stars were identified as variable, while the rest were marked as "not available" (indeterminate identification). This status was mostly assigned to stars for which time series data were not available, making it impossible to determine the period and shape of the light curve. The results are summarized in Table \ref{tab:vari_type}.

The star-forming regions have not yet been widely studied, which limits the possibility of verifying our results. However, there are a few relevant papers that focus on the variability of stars in these regions - specifically in the Pelican Nebula in the direction of IC 5070 \citep{refId01}, in the star-forming region Lynds 1688 \citep{2014AJ....148..122G}  and within the Cygnus OB7 region \citep{Rice_2012}.

In the Pelican Nebula, pre-main-sequence variable stars were classified as candidates for classical T Tauri stars and weak-line T Tauri stars (CTTS, WTTS). Strong periodic changes attributed to modulations caused by cold spots were observed in these objects. For accretion bursts and extinction events, the average amplitudes reached greater than one magnitude in the optical region of the spectrum. 
Both periodic and nonperiodic light curve variations were determined in the Lynds 1688 region, but the authors did not further address the exact type of variability. 
Both of these papers use the Lomb-Scargle method to determine the frequency or period of objects.

T Tauri-type variable stars were also found in Cygnus OB7. In addition, according to them, the variability may be caused by rotationally modulated starspots, while other possible causes include changes in accretion rate, inner hole size, or disk inclination.

Our results show similarities to these studies in determining the type and characteristics of variability, with abnormal object periods and types of variability.

\subsection{Spectral Types}

MKCLASS could determine spectral types in one of the two standard libraries used for about two-thirds of our spectra. When both libraries arrived at a spectral type, they mostly agreed within a margin of error (a few subclasses or luminosity classes).
However, for some stars, they differ a lot. An example is the star 
Gaia DR3 5201536661993294720 classified as kA0hA1mA2 Eu in \texttt{libr18} and 
as M2.5 III with \texttt{libnor36}. Visual inspection agrees more with the 
late-type classification. It also shows hydrogen emission like many of the 
stars in the association, suggesting those are still in the accretion phase of their evolution.
This lends confidence in the classification process.
Generally, following a visual inspection of the spectra, one can say that the spectral 
types determined by \texttt{libnor36} are more likely to be correct than the ones 
by \texttt{libr18}. We conclude that the \texttt{libnor36} library represents 
the characteristics of the spectra for the given signal-to-noise ratio better.

\subsubsection{Possible CP stars}

\begin{figure}
    \centering
    \includegraphics[width = 0.65\textwidth]{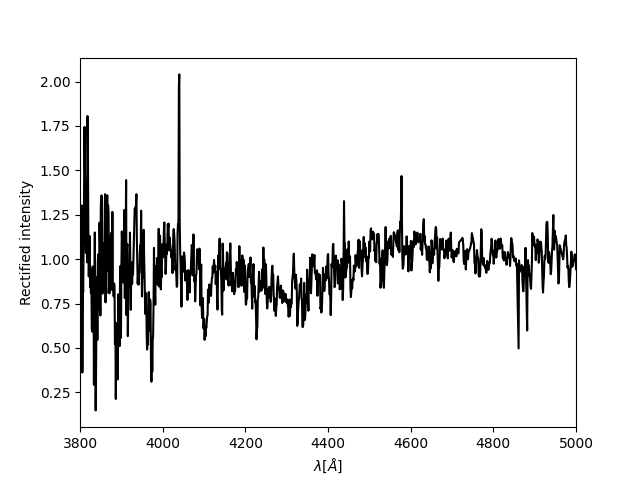}
    \caption{Spectrum of the star that was classified as being peculiar by MKCLASS, 2MASS11101141-7635292/Gaia DR3 5201350019893311744 with spectral type kB7hF6mF7 (\texttt{libnor36}) or kB7hF5mF6 (\texttt{libr18}). }
    \label{fig:cp_member}
\end{figure}

Using MKCLASS, we also detected a few possible CP stars, both as members and non-members. Chemically peculiar stars make up about 10\% of the upper main-sequence
stars (spectral types early B to early F). They are characterized by peculiar
atmospheric abundances that differ significantly from the solar pattern.
Following the classification scheme developed by \cite{1974ARA&A..12..257P}, we have two possible CP1 stars as members, one can be seen in Fig.\ref{fig:cp_member}, of the association and two CP2 stars as non members. 
Additionally, one non-member was classified as Ba or CN star and two members as well
(Table \ref{tab: spectral_types}). It is well known that the unambiguous classification of the different 
CP subgroups is not straightforward \citep{2020A&A...640A..40H}. Therefore, a more 
thorough investigation is needed to confirm the nature of these objects.

\begin{figure}
    \centering
    \includegraphics[width = 0.65\textwidth]{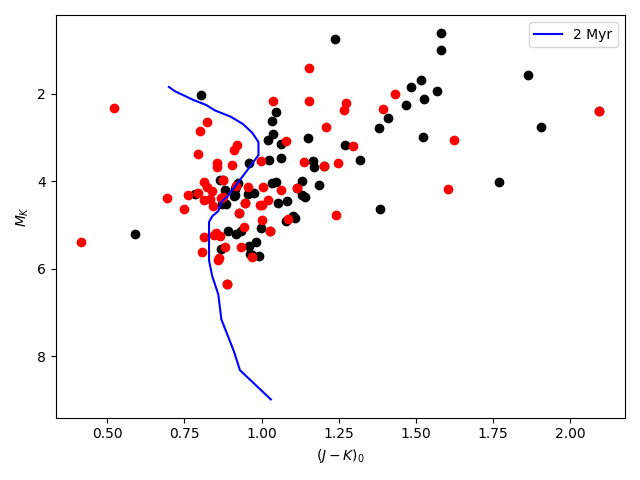}
    \caption{Extinction values were taken from \cite{2007ApJS..173..104L}, the 2 Myr isochrone is from \cite{2015A&A...577A..42B}. Black are all the stars in the association determined by this work, and red are the ones with a light curve.}
    \label{fig:cmd_2mass}
\end{figure}

\section{Conclusions}

In this paper, we studied the light curves in the Chamaeleon I association, where we found 73 members of this association. To determine the spectral type, we used automatic spectral classification with MKCLASS, where we found that most of the Cha I members fall into the cool M-type stars, but we also have a record of a CP star. Our spectra also revealed narrow emission lines associated with ongoing accretion of material onto the star, confirming the young age of these stars.

 The data from which we processed the light and phase curves were taken from the Gaia and NEOWISE surveys and then processed in Python and Peranso software.  Time series data were available for 55 stars from the stellar association from the Gaia archive and 69 of them were available in the NEOWISE archive. 
The phase curves follow an approximately periodic pattern with differences in minimum and maximum amplitudes of about 0.3 mag. Such low amplitudes may be due to the material around the young stars. The phase curves of the studied stars are usually symmetric and not smooth. It may be due to the unstable internal structure of the stars due to its formation, accretion of material onto the star, and stains on its surface. 

\credit{This work was supported by the grant GAČR 23-07605S and was carried out within the institutional support framework 
for the development of the
research organization of Masaryk University.
      Based partly on data acquired at the Anglo-Australian Telescope in the semester 2009A. We acknowledge the traditional custodians of the land on which the AAT stands, the Gamilaraay people, and pay our respects to elders past and present.
      This work presents results from the European Space Agency (ESA) space mission 
      Gaia. Gaia data are being processed by the Gaia Data Processing and Analysis 
      Consortium (DPAC). Funding for the DPAC is provided by national institutions, 
      in particular, the institutions participating in the Gaia Multilateral Agreement 
      (MLA). The Gaia mission website is https://www.cosmos.esa.int/gaia. 
      The Gaia archive website is https://archives.esac.esa.int/gaia.
      This publication makes use of data products from the Two Micron All Sky Survey, which is a joint project of the University of Massachusetts and the Infrared Processing and Analysis Center/California Institute of Technology, funded by the National Aeronautics and Space Administration and the National Science Foundation.
      This research has made use of the WEBDA database, operated at the Department of Theoretical Physics and Astrophysics of Masaryk University.}

\clearpage %%Remove this from your manuscript

\begin{appendix}
\section{Members using HDBSCAN}

\begin{longtable}{c|c|c|c|c|c|c}
\caption{Members determined by the automatic clustering using HDBSCAN.} \label{tab:members_hdbscan} \\
\hline
\textbf{Gaia DR3} & \textbf{2MASS} & \textbf{RA} & \textbf{DEC} & $\pi$ & $\mu_\alpha$ & $\mu_\delta$ \\
    &   & [deg]  & [deg] & [mas] & [mas/yr] & [mas/yr]\\
\hline
\endfirsthead

\multicolumn{7}{c}%
{{\bfseries Table \thetable\ continued from previous page}} \\
\hline
\textbf{Gaia DR3} & \textbf{2MASS} & \textbf{RA} & \textbf{DEC} & $\pi$ & $\mu_\alpha$ & $\mu_\delta$ \\
    &   & [deg]  & [deg] & [mas] & [mas/yr] & [mas/yr]\\
\hline
\endhead

\hline \multicolumn{7}{r}{{Continued on next page}} \\ \hline
\endfoot

\hline \hline
\endlastfoot

5200488346376334848 & 11224794-7728438 & 170.699 & -77.479 & 4.028 & -11.449 & 4.158 \\
5201258622989072640 & 11093777-7710410 & 167.407 & -77.178 & 5.053 & -21.928 & 0.253 \\
5201240309248796672 & 11152180-7724042 & 168.840 & -77.401 & 5.454 & -23.610 & 0.034 \\
5201142688935251968 & 11155827-7729046 & 168.992 & -77.484 & 5.483 & -23.202 & 0.635 \\
5201260237896814080 & 11112260-7705538 & 167.844 & -77.098 & 5.255 & -22.566 & -0.223 \\
5201160525934988288 & 11104959-7717517 & 167.706 & -77.298 & 5.357 & -22.317 & 1.296 \\
5201163480873128064 & 11102852-7716596 & 167.618 & -77.283 & 5.480 & -23.648 & 1.147 \\
5201150973927675904 & 11101153-7733521 & 167.547 & -77.564 & 5.281 & -22.956 & 0.038 \\
5201151180086109696 & 11094918-7731197 & 167.454 & -77.522 & 5.428 & -23.244 & -0.376 \\
5201152245237987072 & 11083952-7734166 & 167.164 & -77.571 & 5.337 & -23.249 & 0.592 \\
5201125272843631616 & 11094525-7740332 & 167.437 & -77.676 & 5.171 & -22.939 & -0.425 \\
5201153619627528960 & 11082238-7730277 & 167.092 & -77.507 & 5.260 & -22.816 & 0.286 \\
5201160663373911552 & 11094260-7725578 & 167.426 & -77.433 & 5.168 & -23.255 & 0.278 \\
5201152760634067712 & 11085421-7732115 & 167.225 & -77.536 & 5.287 & -23.255 & 0.597 \\
5201127918543201664 & 11082410-7741473 & 167.100 & -77.697 & 5.296 & -23.252 & 0.418 \\
5201126063117435136 & 11081648-7744371 & 167.068 & -77.743 & 5.271 & -22.510 & 0.585 \\
5201126441074447872 & 11075809-7742413 & 166.991 & -77.711 & 5.332 & -22.711 & -0.508 \\
5201128055982161280 & 11081896-7739170 & 167.079 & -77.655 & 5.223 & -23.391 & -0.039 \\
5201119878364524928 & 11073832-7747168 & 166.909 & -77.788 & 5.243 & -23.368 & 0.614 \\
5201128743176923520 & 11074610-7740089 & 166.942 & -77.669 & 5.090 & -23.455 & 0.453 \\
5201120634278654208 & 11071148-7746394 & 166.797 & -77.776 & 5.221 & -22.968 & -0.461 \\
5201126750312088320 & 11071330-7743498 & 166.805 & -77.730 & 5.299 & -25.031 & -0.931 \\
5201127334427643008 & 11065733-7742106 & 166.738 & -77.703 & 5.263 & -22.640 & 1.532 \\
5201126990830255872 & 11063799-7743090 & 166.658 & -77.719 & 5.123 & -23.281 & 0.935 \\
5201129292932746880 & 11075225-7736569 & 166.967 & -77.616 & 5.146 & -23.209 & 0.941 \\
5201129705249611008 & 11073519-7734493 & 166.896 & -77.580 & 5.284 & -23.155 & 0.166 \\
5201175987817179136 & 11062877-7737331 & 166.619 & -77.626 & 5.226 & -22.502 & -0.045 \\
5201153791426217472 & 11072040-7729403 & 166.834 & -77.494 & 5.465 & -23.426 & 1.895 \\
5201168944070798848 & 11044258-7741571 & 166.177 & -77.699 & 5.241 & -23.025 & 1.855 \\
5201153172951606784 & 11073686-7733335 & 166.903 & -77.559 & 5.206 & -24.227 & 0.780 \\
5201201139145692800 & 11064346-7726343 & 166.681 & -77.443 & 5.308 & -22.034 & 1.651 \\
5201209351123255296 & 11065906-7718535 & 166.745 & -77.315 & 5.333 & -23.291 & 1.598 \\
5201181313576638208 & 11034186-7726520 & 165.924 & -77.448 & 5.206 & -23.449 & 1.831 \\
5201180248424735488 & 11023265-7729129 & 165.635 & -77.487 & 5.521 & -22.647 & 2.419 \\
5201206361825924992 & 11034764-7719563 & 165.948 & -77.332 & 5.109 & -22.533 & -2.722 \\
5201258863507239424 & 11090512-7709580 & 167.271 & -77.166 & 5.396 & -23.121 & 1.785 \\
5201206052588955264 & 11025504-7721508 & 165.729 & -77.364 & 5.429 & -23.791 & 1.905 \\
5201207736215474688 & 11045701-7715569 & 166.237 & -77.266 & 5.166 & -22.712 & 0.444 \\
5201211172189327744 & 11075993-7715317 & 166.999 & -77.259 & 5.164 & -22.086 & 0.270 \\
5201194816953813504 & 11011370-7722387 & 165.307 & -77.377 & 5.273 & -23.548 & 2.761 \\
5201378641555926784 & 10555973-7724399 & 163.999 & -77.411 & 5.450 & -23.691 & 2.239 \\
5201214367644997376 & 11052272-7709290 & 166.344 & -77.158 & 5.212 & -22.342 & 0.650 \\
5201164202426983040 & 11083905-7716042 & 167.162 & -77.269 & 5.269 & -22.508 & 0.907 \\
5201308101012353664 & 11085464-7702129 & 167.227 & -77.037 & 5.358 & -23.557 & 1.003 \\
5201212649658082560 & 11060466-7710063 & 166.519 & -77.168 & 5.365 & -17.582 & 2.510 \\
5201387776948008320 & 10580597-7711501 & 164.524 & -77.197 & 5.343 & -23.305 & 2.337 \\
5201536661993294720 & 10561638-7630530 & 164.068 & -76.514 & 5.239 & -22.773 & 1.378 \\
5201335481425778560 & 11040425-7639328 & 166.017 & -76.659 & 5.142 & -23.696 & -0.817 \\
5201553704423697792 & 11004022-7619280 & 165.167 & -76.324 & 5.181 & -22.558 & 0.639 \\
5201347064956072448 & 11064180-7635489 & 166.674 & -76.597 & 5.136 & -21.069 & -0.505 \\
5201267625240570240 & 11092913-7659180 & 167.371 & -76.988 & 5.097 & -22.349 & 0.217 \\
5201361599124919936 & 11052472-7626209 & 166.353 & -76.439 & 5.157 & -22.329 & 0.231 \\
5201374209148844672 & 11041060-7612490 & 166.044 & -76.213 & 5.347 & -22.695 & -0.006 \\
5201345484407780608 & 11080234-7640343 & 167.009 & -76.676 & 5.197 & -22.175 & 0.956 \\
5201344659774097664 & 11084069-7636078 & 167.169 & -76.602 & 5.044 & -23.009 & -0.339 \\
5201355444434076416 & 11071181-7625501 & 166.799 & -76.431 & 5.002 & -22.300 & -0.219 \\
5201351085045200640 & 11085497-7632410 & 167.229 & -76.545 & 5.170 & -22.325 & -0.991 \\
5225392628342178688 & 11071915-7603048 & 166.829 & -76.051 & 5.144 & -22.012 & 0.323 \\
5201356337787287552 & 11085090-7625135 & 167.211 & -76.420 & 5.202 & -22.020 & -0.157 \\
5201351772240003456 & 11091380-7628396 & 167.307 & -76.478 & 5.127 & -21.130 & -0.248 \\
5201350049957132800 & 11094621-7634463 & 167.442 & -76.580 & 5.094 & -21.214 & -0.929 \\
5201351428642607744 & 11091812-7630292 & 167.325 & -76.508 & 5.445 & -21.817 & -0.675 \\
5201350157332278784 & 11100369-7633291 & 167.515 & -76.558 & 5.109 & -21.610 & -1.608 \\
5201351527423800320 & 11095407-7629253 & 167.475 & -76.490 & 5.010 & -22.540 & -0.820 \\
5201350019893311744 & 11101141-7635292 & 167.547 & -76.591 & 5.100 & -22.157 & -1.434 \\
5225374074083637376 & 11113965-7620152 & 167.914 & -76.338 & 5.206 & -21.807 & -0.250 \\
5201303325008889856 & 11105333-7634319 & 167.722 & -76.576 & 5.175 & -21.720 & -0.833 \\
5225322603195569920 & 11142906-7625399 & 168.621 & -76.428 & 5.300 & -21.829 & -0.138 \\
5225314975333685248 & 11175211-7629392 & 169.467 & -76.494 & 5.166 & -21.303 & -0.635 \\
5225314356858481280 & 11194214-7623326 & 169.925 & -76.392 & 5.251 & -21.444 & -0.920 \\
5225302777626451328 & 11183379-7643041 & 169.640 & -76.718 & 5.439 & -22.349 & 0.321 \\
5224581944675548032 & 11241186-7630425 & 171.049 & -76.512 & 5.397 & -22.692 & -0.248 \\
5201286110779988480 & 11173792-7646193 & 169.408 & -76.772 & 5.506 & -23.369 & 1.163 \\

\end{longtable}

\begin{landscape}
\section{Stars with light curve}

\begin{longtable} {@{\extracolsep{\fill}}c|c|c|c|c|c|c|c}
\caption{Astrometric properties of the stars in our sample. The sources marked with "Y" in the column "Member" belong to the Cha I association according to our HDBSCAN method.} \label{tab:results_table} \\
\hline
\textbf{Gaia DR3} & \textbf{2MASS} & \textbf{RA} & \textbf{DEC} & $\pi$ & $\mu_{\alpha\star}$ & $\mu_\delta$ & \textbf{Member} \\
& & [deg] & [deg] & [mas] & [mas/yr]& [mas/yr] & \\
\hline
\endfirsthead

\multicolumn{8}{c}%
{{\bfseries Table \thetable\ continued}} \\
\hline
\textbf{Gaia DR3} & \textbf{2MASS} & \textbf{RA} & \textbf{DEC} & $\pi$ & $\mu_{\alpha\star}$ & $\mu_\delta$ & \textbf{Member} \\
& & [deg] & [deg] & [mas] & [mas/yr]& [mas/yr] & \\
\hline
\endhead

\hline \multicolumn{8}{r}{{Continued on next page}} \\ \hline
\endfoot

\hline \hline
\endlastfoot

5201296727939072384 & 11103607-7640412 & 167.650226 & -76.678097 & 0.338500 & -5.130000 & 0.594000 & N \\
5201128743176923520 & 11074610-7740089 & 166.941573 & -77.669144 & 5.089900 & -23.455000 & 0.453000 & Y \\
5201296349981948928 & 11104343-7641132 & 167.679727 & -76.686888 & 3.407400 & -57.715000 & 16.352000 & N \\
5201206052588955264 & 11025504-7721508 & 165.728875 & -77.364076 & 5.428800 & -23.791000 & 1.905000 & Y \\
5201351527423800320 & 11095407-7629253 & 167.474808 & -76.490382 & 5.009900 & -22.540000 & -0.820000 & Y \\
5201351772240003456 & 11091380-7628396 & 167.307163 & -76.477680 & 5.126700 & -21.130000 & -0.248000 & Y \\
5201345484407780608 & 11080234-7640343 & 167.009369 & -76.676203 & 5.196700 & -22.175000 & 0.956000 & Y \\
5201214367644997376 & 11052272-7709290 & 166.344151 & -77.158026 & 5.211600 & -22.342000 & 0.650000 & Y \\
5201520890873415296 & 10592751-7635068 & 164.864719 & -76.585200 & 1.517400 & 1.482000 & 5.467000 & N \\
5201126063117435136 & 11081648-7744371 & 167.068215 & -77.743665 & 5.271500 & -22.510000 & 0.585000 & Y \\
5201160525934988288 & 11104959-7717517 & 167.706150 & -77.297708 & 5.356500 & -22.317000 & 1.296000 & Y \\
5201351428642607744 & 11091812-7630292 & 167.325167 & -76.508130 & 5.445400 & -21.817000 & -0.675000 & Y \\
5201168944070798848 & 11044258-7741571 & 166.177018 & -77.699179 & 5.241900 & -23.025000 & 1.855000 & Y \\
5201163480873128064 & 11102852-7716596 & 167.618441 & -77.283222 & 5.479700 & -23.648000 & 1.147000 & Y \\
5201343693404220800 & 11103368-7639225 & 167.640239 & -76.656235 & 2.815000 & -6.783000 & 0.608000 & N \\
5201258863507239424 & 11090512-7709580 & 167.270889 & -77.166110 & 5.396400 & -23.121000 & 1.785000 & Y \\
5201343693404220800 & 11103368-7639225 & 167.640239 & -76.656235 & 2.815000 & -6.783000 & 0.608000 & N \\
5201206361825924992 & 11034764-7719563 & 165.948150 & -77.332328 & 5.108900 & -22.533000 & -2.722000 & Y \\
5201211172189327744 & 11075993-7715317 & 166.999182 & -77.258813 & 5.163800 & -22.083000 & 0.270000 & Y \\
5201351085045200640 & 11085497-7632410 & 167.228557 & -76.544771 & 5.170200 & -22.325000 & -0.991000 & Y \\
5201356337787287552 & 11085090-7625135 & 167.211703 & -76.420464 & 5.201700 & -22.020000 & -0.158000 & Y \\
5201120634278654208 & 11071148-7746394 & 166.797231 & -77.777590 & 5.221500 & -22.968000 & -0.461000 & Y \\
5201152245237987072 & 11083952-7734166 & 167.164118 & -77.571309 & 5.336600 & -23.249000 & 0.592000 & Y \\
5225374074083637376 & 11113965-7620152 & 167.914869 & -76.337507 & 5.206000 & -21.807000 & -0.250000 & Y \\
5200516693161390208 & 11231140-7713122 & 170.797056 & -77.220033 & 0.765800 & -19.020000 & 0.044000 & N \\
5201119878364524928 & 11073832-7747168 & 166.909111 & -77.787974 & 5.243400 & -23.368000 & 0.614000 & Y \\
5201209351123255296 & 11065906-7718535 & 166.745546 & -77.314816 & 5.332900 & -23.291000 & 1.598000 & Y \\
5201294425836576512 & 11105597-7645325 & 167.732819 & -76.759062 & 6.424700 & -23.748000 & 0.142000 & N \\
5201240309248796672 & 11152180-7724042 & 168.840392 & -77.401069 & 5.454400 & -23.611000 & 0.034000 & Y \\
5201296727939072384 & 11103607-7640412 & 167.650226 & -76.678097 & 0.338500 & -5.130000 & 0.594000 & N \\
5201152760634067712 & 11085421-7732115 & 167.225349 & -77.536565 & 5.286800 & -23.255000 & 0.597000 & Y \\
5201304046563440384 & 11121784-7632352 & 168.074333 & -76.543050 & 0.234600 & -1.046000 & 2.776000 & N \\
5201350049957132800 & 11094621-7634463 & 167.442018 & -76.579586 & 5.094400 & -21.214000 & -0.929000 & Y \\
5225392628342178688 & 11071915-7603048 & 166.829342 & -76.051344 & 5.143500 & -22.012000 & 0.323000 & Y \\
5201335481425778560 & 11040425-7639328 & 166.017285 & -76.659134 & 5.142500 & -23.696000 & -0.817000 & Y \\
5201341945354739840 & 11092482-7642073 & 167.353280 & -76.702055 & 0.381100 & -7.805000 & 2.301000 & N \\
5225314975329893888 & 11175211-7629392 & 169.466782 & -76.494381 & 5.279300 & -22.087000 & -0.591000 & Y \\
5201341567397611008 & 11094586-7643543 & 167.440951 & -76.731825 & 2.308800 & -11.765000 & -0.931000 & N \\
5201129705249611008 & 11073519-7734493 & 166.896085 & -77.580353 & 5.284000 & -23.155000 & 0.166000 & Y \\
5201294425836576512 & 11105597-7645325 & 167.732819 & -76.759062 & 6.424700 & -23.748000 & 0.142000 & N \\
5201151180086109696 & 11094918-7731197 & 167.453908 & -77.522256 & 5.427600 & -23.244000 & -0.376000 & Y \\
5225302777626451328 & 11183379-7643041 & 169.640345 & -76.717789 & 5.439200 & -22.349000 & 0.321000 & Y \\
5201129292932746880 & 11075225-7736569 & 166.967261 & -77.615826 & 5.145600 & -23.209000 & 0.941000 & Y \\
5201175987817179136 & 11062877-7737331 & 166.619369 & -77.625866 & 5.225600 & -22.502000 & -0.045000 & Y \\
5201181313576638208 & 11034186-7726520 & 165.923959 & -77.447773 & 5.206100 & -23.449000 & 1.831000 & Y \\
5201126990830255872 & 11063799-7743090 & 166.657686 & -77.719183 & 5.123500 & -23.281000 & 0.935000 & Y \\
5201536661993294720 & 10561638-7630530 & 164.067798 & -76.514745 & 5.239300 & -22.773000 & 1.378000 & Y \\
5201378641555926784 & 10555973-7724399 & 163.998542 & -77.411075 & 5.450100 & -23.691000 & 2.239000 & Y \\
5201374209148844672 & 11041060-7612490 & 166.043744 & -76.213620 & 5.346900 & -22.695000 & -0.006000 & Y \\
5201128055982161280 & 11081896-7739170 & 167.078596 & -77.654751 & 5.223300 & -23.391000 & -0.039000 & Y \\
5201387776948008320 & 10580597-7711501 & 164.524373 & -77.197222 & 5.343500 & -23.305000 & 2.337000 & Y \\
5201125272843631616 & 11094525-7740332 & 167.437937 & -77.675933 & 5.171300 & -22.940000 & -0.425000 & Y \\
5201153791426217472 & 11072040-7729403 & 166.834516 & -77.494511 & 5.465200 & -23.426000 & 1.895000 & Y \\
5201153172951606784 & 11073686-7733335 & 166.903163 & -77.559270 & 5.205500 & -24.227000 & 0.800000 & Y \\
5201249311500282496 & 11154488-7710433 & 168.936639 & -77.178722 & 0.992500 & -12.646000 & 2.454000 & N \\
5201164202426983040 & 11083905-7716042 & 167.162243 & -77.267816 & 5.268700 & -22.508000 & 0.907000 & Y \\
5201260237896814080 & 11112260-7705538 & 167.843905 & -77.098305 & 5.255000 & -22.566000 & -0.223000 & Y \\
5201311021590093056 & 11053220-7659319 & 166.384369 & -76.992148 & 1.233400 & -4.065000 & -2.136000 & N \\
5201126441074447872 & 11075809-7742413 & 166.991460 & -77.711498 & 5.332900 & -22.711000 & -0.508000 & Y \\
5201160663373911552 & 11094260-7725578 & 167.426787 & -77.432795 & 5.168300 & -23.255000 & 0.278000 & Y \\
5201347064956072448 & 11064180-7635489 & 166.673756 & -76.596956 & 5.136400 & -21.069000 & -0.505000 & Y \\
5201142688935251968 & 11155827-7729046 & 168.992376 & -77.484582 & 5.482800 & -23.202000 & 0.635000 & Y \\
5224560607277999232 & 11224520-7635119 & 170.688236 & -76.586628 & 0.248800 & -6.746000 & 3.422000 & N \\
5201314045247097472 & 11052179-7652400 & 166.340703 & -76.877770 & 1.210800 & -11.226000 & 4.203000 & N \\
5201127334427643008 & 11065733-7742106 & 166.738330 & -77.702939 & 5.263500 & -22.641000 & 1.532000 & Y \\
5201180248424735488 & 11023265-7729129 & 165.635660 & -77.486888 & 5.521300 & -22.647000 & 2.419000 & N \\
5201127918543201664 & 11082410-7741473 & 167.099990 & -77.696509 & 5.296100 & -23.253000 & 0.418000 & Y \\
5201126750312088320 & 11071330-7743498 & 166.804806 & -77.730482 & 5.299000 & -25.031000 & -0.931000 & Y \\
5224561844228438016 & 11205545-7632234 & 170.232604 & -76.539848 & 0.901100 & -5.448000 & 3.232000 & N \\
5201355444434076416 & 11071181-7625501 & 166.798760 & -76.430557 & 5.002300 & -22.300000 & -0.220000 & Y \\
5201212649658082560 & 11060466-7710063 & 166.519010 & -77.168414 & 5.365000 & -17.582000 & 2.510000 & Y \\
5225322603195569920 & 11142906-7625399 & 168.620648 & -76.427806 & 5.300000 & -21.829000 & -0.138000 & Y \\
5225314356858481280 & 11194214-7623326 & 169.925137 & -76.392365 & 5.250900 & -21.444000 & -0.920000 & Y \\
5200523255870845696 & 11225375-7707140 & 170.723897 & -77.120527 & 0.621500 & -3.641000 & 0.934000 & N \\
5224581944675548032 & 11241186-7630425 & 171.048981 & -76.511838 & 5.396600 & -22.692000 & -0.248000 & Y \\
5201350019893311744 & 11101141-7635292 & 167.547056 & -76.591463 & 5.100500 & -22.157000 & -1.434000 & Y \\
5201308101012353664 & 11085464-7702129 & 167.227151 & -77.036937 & 5.357500 & -23.557000 & 1.003000 & Y \\
5201201139145692800 & 11064346-7726343 & 166.680706 & -77.442878 & 5.308500 & -22.034000 & 1.651000 & Y \\
5201153619627528960 & 11082238-7730277 & 167.092735 & -77.507708 & 5.260100 & -22.816000 & 0.286000 & Y \\
5201194816953813504 & 11011370-7722387 & 165.306647 & -77.377388 & 5.273200 & -23.548000 & 2.761000 & Y \\
5201303325008889856 & 11105333-7634319 & 167.721768 & -76.575572 & 5.175300 & -21.720000 & -0.833000 & Y \\
5201301778820649088 & 11121969-7639320 & 168.081772 & -76.658828 & 1.921600 & -17.962000 & 4.771000 & N \\
5224581944675548032 & 11241186-7630425 & 171.048981 & -76.511838 & 5.396600 & -22.692000 & -0.248000 & Y \\
5201286110779988480 & 11173792-7646193 & 169.407586 & -76.772032 & 5.506500 & -23.369000 & 1.163000 & Y \\
5201350157332278784 & 11100369-7633291 & 167.514927 & -76.558118 & 5.108500 & -21.610000 & -1.608000 & Y \\
5201361599124919936 & 11052472-7626209 & 166.352595 & -76.439098 & 5.156500 & -22.329000 & 0.231000 & Y \\
5201303870467508736 & 11121409-7635003 & 168.058536 & -76.583373 & 0.158100 & -6.247000 & 4.998000 & N \\
5201553704423697792 & 11004022-7619280 & 165.167214 & -76.324432 & 5.181100 & -22.558000 & 0.639000 & Y \\
5201341155080740480 & 11095146-7645452 & 167.464269 & -76.762580 & 0.792800 & -6.677000 & 3.829000 & N \\
5201267625240570240 & 11092913-7659180 & 167.370830 & -76.988365 & 5.097400 & -22.349000 & 0.217000 & Y \\
5201344659774097664 & 11084069-7636078 & 167.169171 & -76.602173 & 5.044500 & -23.009000 & -0.339000 & Y \\
5201150973927675904 & 11101153-7733521 & 167.547471 & -77.564459 & 5.281800 & -22.956000 & 0.038000 & Y \\
5201207736215474688 & 11045701-7715569 & 166.237691 & -77.265867 & 5.165600 & -22.712000 & 0.444000 & Y \\

\end{longtable}
\end{landscape}

\clearpage

\section{Spectral Types}

\begin{longtable}{c|c|c|c|c}
\caption{Spectral types of the stars with a light curve}
\label{tab: spectral_types} \\
\hline 
\textbf{Gaia DR3} & \textbf{2MASS} & \textbf{libr18} & \textbf{libnor36} & \textbf{Member} \\
\hline
\endfirsthead

\multicolumn{5}{c}%
{{\bfseries Table \thetable\ continued }} \\
\hline
\textbf{Gaia DR3} & \textbf{2MASS} & \textbf{libr18} & \textbf{libnor36} & \textbf{Member} \\
\hline
\endhead
\hline
\multicolumn{5}{r}{{Continued on next page}} \\
\hline
\endfoot

\hline \hline
\endlastfoot

5201296727939072384 & 11103607-7640412 &  ?? &  K2 IV metal-weak  & N \\
5201128743176923520 & 11074610-7740089 &  Unclassifiable              &  Unclassifiable              & Y \\
5201206052588955264 & 11025504-7721508 &  ?      &  Unclassifiable              & Y \\
5201351527423800320 & 11095407-7629253 &  emission-line? &  emission-line? & Y \\
5201351772240003456 & 11091380-7628396 &  Unclassifiable              &  M5 III   & Y \\
5201345484407780608 & 11080234-7640343 &  ?      &  M3 III   & Y \\
5201214367644997376 & 11052272-7709290 &  ?? &  emission-line? & Y \\
5201520890873415296 & 10592751-7635068 &  F9 III-IV Fe-0.6   &  G0 IV Fe-0.6   & N \\
5201126063117435136 & 11081648-7744371 &  Unclassifiable              &  M4 IV-V   & Y \\
5201160525934988288 & 11104959-7717517 &  emission-line? &  emission-line? & Y \\
5201351428642607744 & 11091812-7630292 &  ?      &  Unclassifiable              & Y \\
5201168944070798848 & 11044258-7741571 &  ?      &  M3.5 IV   & Y \\
5201163480873128064 & 11102852-7716596 &  Unclassifiable              &  M4 III   & Y \\
5201343693404220800 & 11103368-7639225 &  emission-line? &  emission-line? & N \\
5201258863507239424 & 11090512-7709580 &  Unclassifiable              &  M4 III   & Y \\
5201206361825924992 & 11034764-7719563 &  kB8hA6mF6   &  Unclassifiable              & Y \\
5201211172189327744 & 11075993-7715317 &  Unclassifiable              &  Unclassifiable              & Y \\
5201351085045200640 & 11085497-7632410 &  ?      &  Unclassifiable              & Y \\
5201356337787287552 & 11085090-7625135 &  Unclassifiable              &  M4 III   & Y \\
5201120634278654208 & 11071148-7746394 &  Unclassifiable              &  M2.5 V   & Y \\
5201152245237987072 & 11083952-7734166 &  G7 III-IV Fe-1.3   &  Unclassifiable              & Y \\
5225374074083637376 & 11113965-7620152 &  emission-line? &  emission-line? & Y \\
5200516693161390208 & 11231140-7713122 &  F6 V  &  F6 V  & N \\
5201119878364524928 & 11073832-7747168 &  Unclassifiable              &  M5 III   & Y \\
5201209351123255296 & 11065906-7718535 &  Unclassifiable              &  M5 III   & Y \\
5201294425836576512 & 11105597-7645325 &  emission-line? &  emission-line? & N \\
5201240309248796672 & 11152180-7724042 &  Unclassifiable              &  M4 III-IV   & Y \\
5201152760634067712 & 11085421-7732115 &  G3 Ib-II   &  Unclassifiable              & Y \\
5201304046563440384 & 11121784-7632352 &  K2 III-IV   &  K2 III-IV   & N \\
5201350049957132800 & 11094621-7634463 &  Unclassifiable              &  Unclassifiable              & Y \\
5225392628342178688 & 11071915-7603048 &  Unclassifiable              &  M2 V   & Y \\
5201335481425778560 & 11040425-7639328 &  Unclassifiable              &  M0 Ia   & Y \\
5201341945354739840 & 11092482-7642073 &  K3 II  CN1  &  K4 III CN2  & N \\
5201341567397611008 & 11094586-7643543 &  kA2hA4mA7   &  kA1hA3mA7  Eu & N \\
5201129705249611008 & 11073519-7734493 &  Unclassifiable              &  M3.5 V   & Y \\
5201151180086109696 & 11094918-7731197 &  Unclassifiable              &  Unclassifiable              & Y \\
5225302777626451328 & 11183379-7643041 &  G9 V Fe-2.4   &  M5 IV-V   & Y \\
5201129292932746880 & 11075225-7736569 &  kA3hF5mK0   &  M3.5 V   & Y \\
5201175987817179136 & 11062877-7737331 &  Unclassifiable              &  Unclassifiable              & Y \\
5201181313576638208 & 11034186-7726520 &  kB8hA1mA7  Eu &  Unclassifiable              & Y \\
5201126990830255872 & 11063799-7743090 &  K2 V   &  M5 III   & Y \\
5201536661993294720 & 10561638-7630530 &  kA0hA1mA2  Eu &  M2.5 III   & Y \\
5201378641555926784 & 10555973-7724399 &  emission-line? &  M2 III   & Y \\
5201374209148844672 & 11041060-7612490 &  F9 V  &  M5 III   & Y \\
5201128055982161280 & 11081896-7739170 &  A0 V  SrEu  &  M3.5 III   & Y \\
5201387776948008320 & 10580597-7711501 &  Unclassifiable              &  K5 0  Ba  & Y \\
5201125272843631616 & 11094525-7740332 &  Unclassifiable              &  Unclassifiable              & Y \\
5201153791426217472 & 11072040-7729403 &  Unclassifiable              &  M4.5 IV   & Y \\
5201153172951606784 & 11073686-7733335 &  F6 V  Sr  &  Unclassifiable              & Y \\
5201249311500282496 & 11154488-7710433 &  F7 III-IV  &  F8 IV  & N \\
5201164202426983040 & 11083905-7716042 &  K2 V  Ba  &  K2 IV-V  CN1 CH-2 Ba  & Y \\
5201260237896814080 & 11112260-7705538 &  K2 V   &  M5 IV-V   & Y \\
5201311021590093056 & 11053220-7659319 &  G0 IV  &  G1 V  & N \\
5201126441074447872 & 11075809-7742413 &  Unclassifiable              &  Unclassifiable              & Y \\
5201160663373911552 & 11094260-7725578 &  ?? &  Unclassifiable              & Y \\
5201347064956072448 & 11064180-7635489 &  emission-line? &  emission-line? & Y \\
5201142688935251968 & 11155827-7729046 &  G9 III Fe-2.3   &  M4.5 IV   & Y \\
5224560607277999232 & 11224520-7635119 &  G7 III  &  G9 IV  & N \\
5201314045247097472 & 11052179-7652400 &  emission-line? &  F8 V  & N \\
5201127334427643008 & 11065733-7742106 &  Unclassifiable              &  M4 III   & Y \\
5201180248424735488 & 11023265-7729129 &  Unclassifiable              &  M5 III   & N \\
5201127918543201664 & 11082410-7741473 &  A6 mA0 V Lam Boo  &  M2.5 III   & Y \\
5201126750312088320 & 11071330-7743498 &  Unclassifiable              &  Unclassifiable              & Y \\
5224561844228438016 & 11205545-7632234 &  F8 III-IV  &  F9 IV-V  & N \\
5201355444434076416 & 11071181-7625501 &  Unclassifiable              &  M5 III   & Y \\
5201212649658082560 & 11060466-7710063 &  Unclassifiable              &  M3.5 V   & Y \\
5225322603195569920 & 11142906-7625399 &  ?      &  M4.5 IV-V   & Y \\
5225314356858481280 & 11194214-7623326 &  G9 V Fe-3.1   &  M4.5 IV-V   & Y \\
5200523255870845696 & 11225375-7707140 &  F7 IV  &  emission-line? & N \\
5224581944675548032 & 11241186-7630425 &  ?? &  M2 III   & Y \\
5201350019893311744 & 11101141-7635292 &  kB7hF5mF6   &  kB7hF6mF7   & Y \\
5201308101012353664 & 11085464-7702129 &  emission-line? &  emission-line? & Y \\
5201201139145692800 & 11064346-7726343 &  ?      &  M0 IV-V   & Y \\
5201153619627528960 & 11082238-7730277 &  Unclassifiable              &  Unclassifiable              & Y \\
5201194816953813504 & 11011370-7722387 &  Unclassifiable              &  Unclassifiable              & Y \\
5201303325008889856 & 11105333-7634319 &  emission-line? &  emission-line? & Y \\
5201301778820649088 & 11121969-7639320 &  K1 V   &  K3 IV-V   & N \\
5201286110779988480 & 11173792-7646193 &  emission-line? &  emission-line? & Y \\
5201350157332278784 & 11100369-7633291 &  F6 V Fe-0.9   &  G0 V Fe-1.2   & Y \\
5201361599124919936 & 11052472-7626209 &  emission-line? &  M2 V   & Y \\
5201303870467508736 & 11121409-7635003 &  G4 II-III  &  G2 V  & N \\
5201553704423697792 & 11004022-7619280 &  emission-line? &  emission-line? & Y \\
5201341155080740480 & 11095146-7645452 &  F7 II  CN2 Sr  &  F9 V  Sr  & N \\
5201267625240570240 & 11092913-7659180 &  Unclassifiable              &  M0 III-IV   & Y \\
5201344659774097664 & 11084069-7636078 &  emission-line? &  M2.5 V   & Y \\
5201150973927675904 & 11101153-7733521 &  Unclassifiable              &  M4 III   & Y \\
5201207736215474688 & 11045701-7715569 &  ?      &  K7 V   & Y \\

\setlength{\tabcolsep}{4pt}
\end{longtable}

\newpage

\begin{longtable}{c|c|c|c|c|c}
\caption{VSX determination of variability type} \label{tab:vari_type} \\
\hline
\textbf{Gaia DR3} & \textbf{RA} & \textbf{DEC} &  \textbf{Type} & \textbf{P} & \textbf{Variability flag} \\
    &   [deg]  & [deg] & & [days]  &  \\ 
\hline
\endfirsthead

\multicolumn{5}{c}%
{{\bfseries Table \thetable\ continued }} \\
\hline
\textbf{Gaia DR3} & \textbf{RA} & \textbf{DEC} &  \textbf{Type} & \textbf{P} & \textbf{Variability flag} \\
    &   [deg]  & [deg] & [days] & & \\
\hline
\endhead
\hline
\multicolumn{5}{r}{{Continued on next page}} \\
\hline
\endfoot

\hline \hline
\endlastfoot
5201351772240003456 & 167.3072 & -76.4777 & ROT  & 0.6759     & variable       \\
5201160525934988288 & 167.7062 & -77.2977 & INT  & 0.7406     & variable       \\
5201351428642607744 & 167.3252 & -76.5081 & TTS  & 6.7818     & not available  \\
5201206361825924992 & 165.9482 & -77.3323 & INT  & 6.2562     & variable       \\
5201351085045200640 & 167.2286 & -76.5448 & INT  & 6.6733     & variable       \\
5201152245237987072 & 167.1641 & -77.5713 & INT  & 0.5708     & variable       \\
5225374074083637376 & 167.9149 & -76.3375 & CTTS & 30.5164    & variable       \\
5201209351123255296 & 166.7455 & -77.3148 & INS  & 0.6484     & variable       \\
5201152760634067712 & 167.2253 & -77.5366 & INT  & 4.3221     & variable       \\
5201129705249611008 & 166.8961 & -77.5804 & INT  & 1.6145     & variable       \\
5201129292932746880 & 166.9673 & -77.6158 & INT  & 4.8311     & variable       \\
5201181313576638208 & 165.924  & -77.4478 & EB   & 1.6314     & variable       \\
5201378641555926784 & 163.9985 & -77.4111 & INSB & 0.8509     & variable       \\
5201153172951606784 & 166.9032 & -77.5593 & INT  & 1.2339     & variable       \\
5201347064956072448 & 166.6738 & -76.597  & CTTS & 1.055      & variable       \\
5201350019893311744 & 167.5471 & -76.5915 & CTTS & 5.1704     & variable       \\
5201308101012353664 & 167.2272 & -77.0369 & INS  & 0.6672     & variable       \\
5201303325008889856 & 167.7218 & -76.5756 & INSB & 1.1688     & variable       \\
5201207736215474688 & 166.2377 & -77.2659 & SR   & 1.4299     & variable       \\
5201128743176923520 & 166.9416 & -77.6691 & INT  & -          & not available  \\
5201351527423800320 & 167.4748 & -76.4904 & INT  & -          & not available  \\
5201350049957132800 & 167.4420 & -76.5796 & INB  & -          & variable       \\
5201378641555926784 & 163.9985 & -77.4111 & INSB & -          & not available  \\
5201153172951606784 & 166.9032 & -77.5593 & INT  & -          & not available  \\
5201164202426983040 & 167.1622 & -77.2678 & INT  & -          & not available  \\
5201308101012353664 & 167.2272 & -77.0369 & INS  & -          & not available  \\
5201303325008889856 & 167.7218 & -76.5756 & INSB & -          & not available  \\
5201350157332278784 & 167.5149 & -76.5581 & INT  & -          & not available  \\
5201303870467508736 & 168.0585 & -76.5834 & CTTS & -          & not available  \\

\end{longtable}
\noindent \textbf{Note 1:} INT: Orion variables of the T Tauri-type, ROT: Spotted stars that were not classified into a specific class, TTS: T Tauri Stars, CTTS: Classical T Tauri stars, INS: Orion variables showing rapid light variations, EB: eclipsing binary, INB: Orion variable type of intermediate and late spectral types, INSB: Rapid irregular variables of the intermediate and late spectral types observed in nebulosity, SR: Semi-regular variables. \citep{vsx-type} 

\noindent \textbf{Note 2:} Variability flag: variability status according to Gaia DR3 catalogue.
\end{appendix}

\printcredits

\bibliographystyle{cas-model2-names}
% \bibliographystyle{elsarticle-harv}
% \bibliographystyle{plainnat}

\bibliography{cha_vars.bib}
%\printbibliography



\end{document}